\documentclass[preprint2]{aastex}
\newcommand\samename{\vrule height0.4pt depth0.0pt width1.0in \thinspace.}
\shorttitle{Maximum young cluster mass in M~33}
\shortauthors{Gonz\'alez-L\'opezlira et al.}

\begin{document}

\title{Gas surface density, star formation rate surface density, and the maximum mass of young
star clusters in a disk galaxy. I. The flocculent galaxy M~33}

\author{Rosa A.\ Gonz\'alez-L\'opezlira\altaffilmark{1,2}, 
Jan Pflamm-Altenburg\altaffilmark{1}, \& Pavel Kroupa\altaffilmark{1}}

\altaffiltext{1}{Argelander Institut f\"ur Astronomie, Universit\"at Bonn, Auf dem H\"ugel 71, D-53121 Bonn, Germany}
\altaffiltext{2}{On sabbatical leave from the Centro de Radioastronom\'{\i}a y Astrof\'{\i}sica, UNAM, Campus Morelia, Michoac\'an, M\'exico, C.P. 58089; 
{\tt r.gonzalez@crya.unam.mx}}

\begin{abstract}
We analyze the relationship between maximum cluster mass, $M_{\rm max}$, and 
surface densities of total gas ($\Sigma_{\rm gas}$), molecular
gas ($\Sigma_{\rm H_2}$) and star formation rate ($\Sigma_{\rm SFR}$)
in the flocculent galaxy M~33, using published gas data and
a catalog of more than 600 young star clusters in its disk. 
By comparing the radial distributions of gas and
most massive cluster masses, we find that 
$M_{\rm max} \propto \Sigma_{\rm gas}^{4.7 \pm 0.4}$, 
$M_{\rm max} \propto \Sigma_{\rm H_2}^{1.3 \pm 0.1}$, 
and $M_{\rm max} \propto \Sigma_{\rm SFR}^{1.0 \pm 0.1}$.
We rule out that these correlations 
result from the size of sample; hence,  
the change of the maximum cluster mass must be due to physical causes.
\end{abstract}


\section{Introduction}

A long-standing problem of galaxy formation and
evolution has been understanding the
relation between gas surface density ($\Sigma_{\rm gas}$) and star formation rate
(SFR; i.e., the star formation law). 
In the last half century, significant efforts
have been undertaken to clarify this subject. Empirical
correlations have been found between the disk surface density
of star formation, $\Sigma_{\rm SFR}$,
and the gas surface density, $\Sigma_{\rm gas}$ --either total, neutral or
molecular (usually based on CO). Examples of these are
$\Sigma_{\rm SFR} = A\Sigma_{\rm gas}^N$
\citep{schm59,schm63,kenn98},
and $\Sigma_{\rm SFR} = A'\Sigma_{\rm gas}\Omega_{\rm gas}$,
with $\Omega_{\rm gas}$ the average angular velocity of the gas in the disk
\citep[e.g.,][]{silk97,kenn98}.
Data are often compatible with more than one correlation
at a time, a situation that does not help to clarify what processes really drive star formation rates
(e.g., turbulence, large scale shocks, gravitational instabilities,
shear, pressure).

The SFR is often measured through the emission in the H$\alpha$ line,
that is due to the reprocessed ionizing photons produced by O or early B-type stars, 
or via the non-ionizing far ultraviolet (FUV) flux dominated 
by B-type stars.
In order to infer the SFR in this way, then, the relation between the mass
in massive stars and the rest, i.e., the
initial mass function (IMF), has to be known.\footnote{The assumption
also has to be made that the star formation activity is constant 
during the lifetime of the massive stars ($\approx 10^7$ yr for O-stars,
and $\approx 10^8$ yr, for B-stars).} 

Different workers \citep[e.g.,][]{sull00,bell01,
meur09,lee09,bose09} have found that, on average, 
for large samples of galaxies,
the SFRs inferred, respectively, from the H$\alpha$ line and the FUV flux 
are not consistent with a universal IMF. Instead,
the ratio between H$\alpha$ and FUV seems to decline with 
galaxy luminosity or galaxy
mass or SFR or SFR per area. 
\citet{lee09} conclude that  
none of the following factors can cause this
trend if acting alone: uncertainties in stellar evolution and atmospheres, 
effects of different metallicities, variations of
star formation histories, photon leakage, extinction,
and stochasticity of massive star formation.

One particular case of the mismatch between the H$\alpha$ and the
FUV emissions occurs in the outer regions of disk galaxies.
Historically, the H$\alpha$ cut-off there led to the concept
of a gas surface density threshold, below which 
star formation activity was inhibited, and the relationships
shown above between $\Sigma_{\rm gas}$ and SFR broke down \citep{kenn98}. 
Recently available UV data, however, show that there is no   
corresponding ``FUV cut-off", and that star formation
also goes on in outer spiral disks \citep{bois07}.

\citet{pfla08} propose the {\em local} 
integrated galaxial IMF [(L)IGIMF] theory, that is able to explain 
both the H$\alpha$ cut-off in galaxy disks, and the trend in 
the ratio of H$\alpha$ to FUV fluxes with galaxy mass. 
The (L)IGIMF theory deals with surface densities 
(e.g., the local IMF and SFR densities), and 
is an outgrowth of the IGIMF theory 
\citep{krou03,weid05,pfla09}. The IGIMF theory is based on the simple  
concept that star formation,
in the form of embedded clusters, occurs
in molecular cloud cores \citep[e.g.,][]{lada03}.
Although most of these embedded clusters do not
survive as bound star clusters the expulsion of their residual gas,
the determination of the stellar IMF of the whole galaxy 
or of a part of it at a given time
reduces to adding the stellar IMFs of all newly formed embedded clusters.
On the other hand, the embedded cluster maximum mass,
$M_{\rm ecl,max}$, seems to be a function of the total SFR
\citep{weid04a}, and the stellar upper mass limit of each
cluster's IMF is a function of the total star cluster mass
\citep[e.g.,][]{elme83,weid04b,weid06,weid10}.
The resulting IGIMF, then, depends on the
SFR, which itself depends on the gas density.
In order to account for the H$\alpha$ cut-off in exponential gas disks, 
\citet{pfla08} adopt the ansatz $M_{\rm ecl,max}\propto \Sigma_{\rm gas}^{3/2}$. 

Another correlation between maximum cluster mass and SFR surface
density has been proposed by \citet{bill02}, based
on the star formation law, and assuming 
the ambient interstellar medium and the cluster-forming
cloud cores are in pressure equilibrium. 
In this scenario, $M_{\rm ecl,max} 
\propto \Sigma_{\rm SFR}^\eta$, with $2/3 \leq \eta \leq 2$.
The first value is expected in the case of equal (volume) density
clusters \citep{bill02}, while the second would occur if clusters have equal sizes \citep{lars02},
as is supported by the very weak birth radius-cluster mass relation \citep{mark12}.

In view of the limited availability of sufficiently 
accurate cluster mass determinations a decade ago, the relationship 
between maximum cluster {\em luminosity} and $\Sigma_{\rm SFR}$ 
was investigated by \citet{lars02} using {\sl HST} 
data of 6 spiral galaxies. He concluded, however, that the effects of random sampling
statistics in determining the brightest observed cluster luminosities would
make it hard to unveil the connection between physical 
processes and maximum star cluster mass, and that a study performed  
with masses, not luminosities, would be much better to establish
how maximum cluster mass depends on galaxy properties. 
At present, the widely held understanding is that the most massive object 
in an ensemble of clusters scales with the sample size as a result of
statistical variations. On the other hand, if there is a physical link 
between the most massive clusters and the SFR, then a potentially very 
powerful new method of measuring star formation histories of
galaxies is opened up \citep{masc07}.

Here, we test the environment dependent IGIMF ansatz versus the 
stochastic sampling ansatz. We make a direct comparison between cluster
mass and gas surface density in M~33, using data from the literature,
with the aim of investigating whether $\Sigma_{\rm gas}$ 
has a role in determining maximum cluster mass. We will from now on
refer to maximum cluster mass as $M_{\rm max}$ exclusively. 

\section{Star cluster data.}

\citet{shar11} have recently published a study of 
648 clusters younger than 10$^8$ yr in
the disk of M~33, out to $\approx$ 16 kpc of the galaxy center, 
selected from the Spitzer 24 $\mu$m image of the
galaxy \citep{verl07}. \citeauthor*{shar11} obtain galactocentric radii for
the clusters using the warped disk model by \citet{corb97}.
They derive ages, masses, and extinction corrections from the
comparison between cluster spectral energy distributions (SEDs)
and Starbust99 \citep{leit99,vazq05} models; \citeauthor{shar11} 
assume a distance to M~33 of 840 kpc \citep{free91}.\footnote{ 
At this distance, 1$^{\prime\prime} =$ 4.1 pc; $R_{25}$, the galactocentric distance of
the isophote with surface brightness in the $B$-band $\mu_{B} = 25$ mag $\sq^{\prime\prime -1}$,
is 8.6 kpc or $35\farcm4$.}
The SEDs are built with aperture
photometry based on the Galaxy 
Evolution Explorer (GALEX) satellite far- and near-UV data \citep{gild07}; 
an H$\alpha$ map \citep{gree98,hoop00}; and mid- and far-IR images 
\citep[8, 24, 70, and 160 $\mu$m;][]{verl07}. 
\citet{shar11} compare the masses and the 24 $\mu$m semi-major axes $r$ of their clusters; they 
find a large scatter for $r <$ 10 pc (about 13\% of  their sample), 
and $M_{\rm clust} \propto r^{2.09 \pm 0.01}$ for clusters with $r >$ 10 pc. 

We extracted the masses and galactocentric radii of 
the clusters from Fig.\ 13 in \citet{shar11} using the tool Dexter
\citep{deml01}. 
Owing to different surface brightness and crowding characteristics, the completeness limit 
varies with radius, and goes from 
800-1000 $M_\odot$ within about 0.6 $R_{25}$ to $\approx 300 M_\odot$
beyond $R_{25}$. In what follows, we will restrict ourselves to
working with the 258 clusters with at least $10^3 M_\odot$.

\section{ISM data.}

\citet{corb03} and \citet{heye04} publish, respectively,
neutral and  molecular gas radial profiles of M~33, assuming
a distance of 840 kpc and a location of the galaxy center
at RA (J2000) = 01:33:50.89 and Dec(J2000) =
30:39:36.7. In both works, the data are deprojected with the model of
\citet{corb97}. While the published HI profile extends as far as 1.5 $R_{25}$,
the H$_2$ one reaches only to 0.8 $R_{25}$.

For the CO to H$_2$ conversion, \citeauthor{heye04} take a constant
factor $X = 3\times 10^{20}$ cm$^{-2}$ (K km s$^{-1}$)$^{-1}$.\footnote{
$\Sigma_{\rm H_2} = 2~m_{\rm H}N$(H$_2$); the molecular hydrogen column density is 
$N$(H$_2$) = $X \int T_{\rm mb}(\rm CO (1-0))$d$v$, 
where $T_{\rm mb}$ is the main beam brightness temperature.} 
\citet{heye04} also derive a radial profile of $\Sigma_{\rm SFR}$ from 
the FIR luminosity, using IRAS HiRes 60 and 100 $\mu$m images
of M~33.\footnote{
$\Sigma_{\rm SFR} = 3.8 \times 10^{-16}(2.58 \langle I_{60} \rangle_\Omega + 
\langle I_{100} \rangle_\Omega)M_\odot {\rm pc^{-2} yr^{-1}}$, 
with $\langle I_{60} \rangle_\Omega$ and 
$\langle I_{100} \rangle_\Omega$ the mean 60 and 100 $\mu$m intensities,
respectively, within a solid angle $\Omega$ .
} 
They find $\Sigma_{\rm SFR} = (3.2 \pm 0.2)~\Sigma^{1.36 \pm 0.08}_{\rm gas,molecular}$,
but $\Sigma_{\rm SFR} = (0.0035 \pm 0.066)~\Sigma^{3.3 \pm 0.07}_{\rm gas,total}$.   
They also ascribe the steep slope of the correlation with total gas to the very
shallow radial distribution of atomic gas.  
Once again, we use Dexter \citep{deml01}
to extract the profiles; we calculate total gas mass surface density
as $\Sigma_{\rm gas} = 1.36(\Sigma_{\rm H_2} + \Sigma_{\rm HI})$,
in order to include helium.

\section{Analysis and discussion}
	
\begin{figure*}
\includegraphics[angle=0.,width=0.5\hsize,clip=]{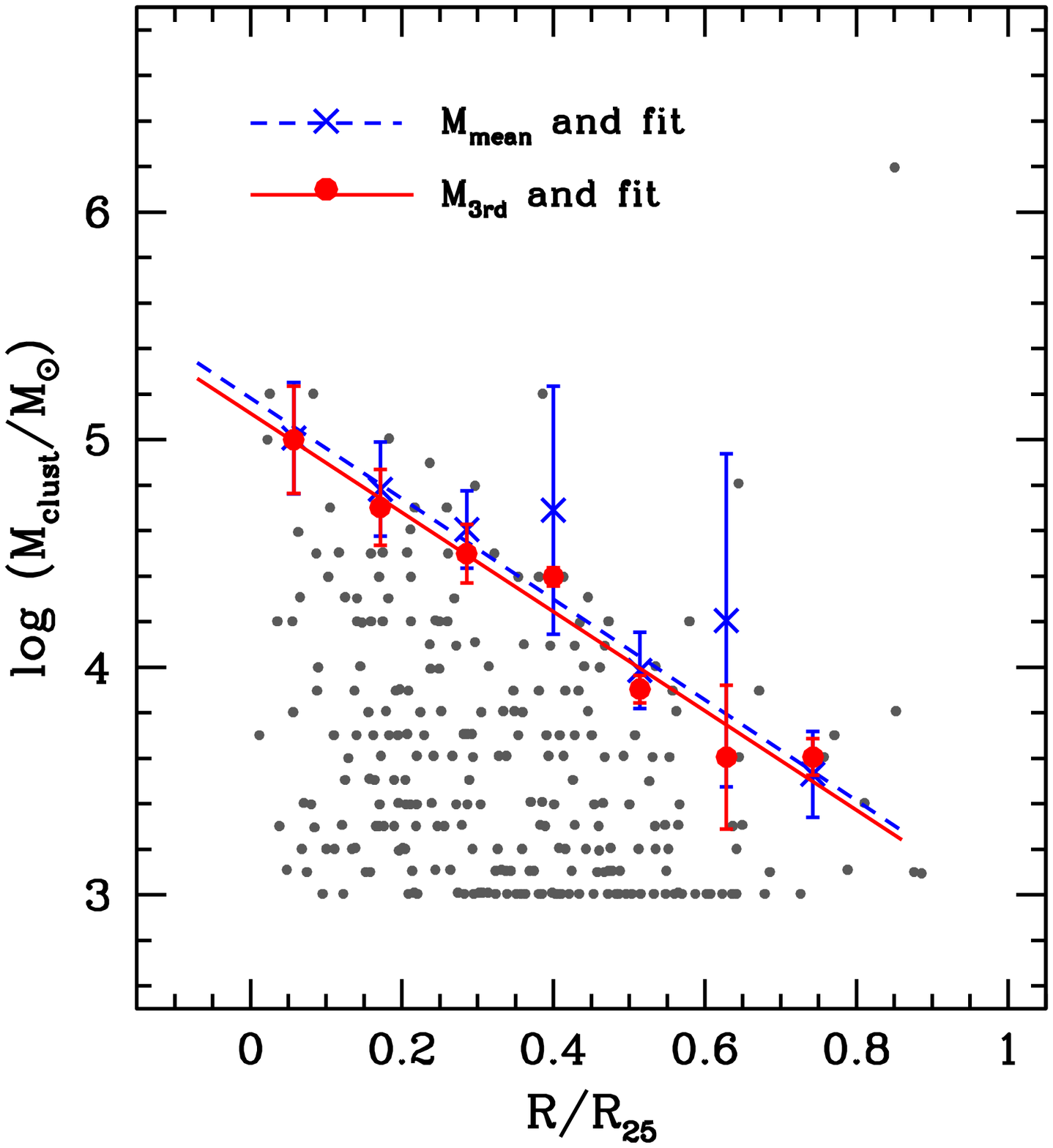}
\includegraphics[angle=0.,width=0.5\hsize,clip=]{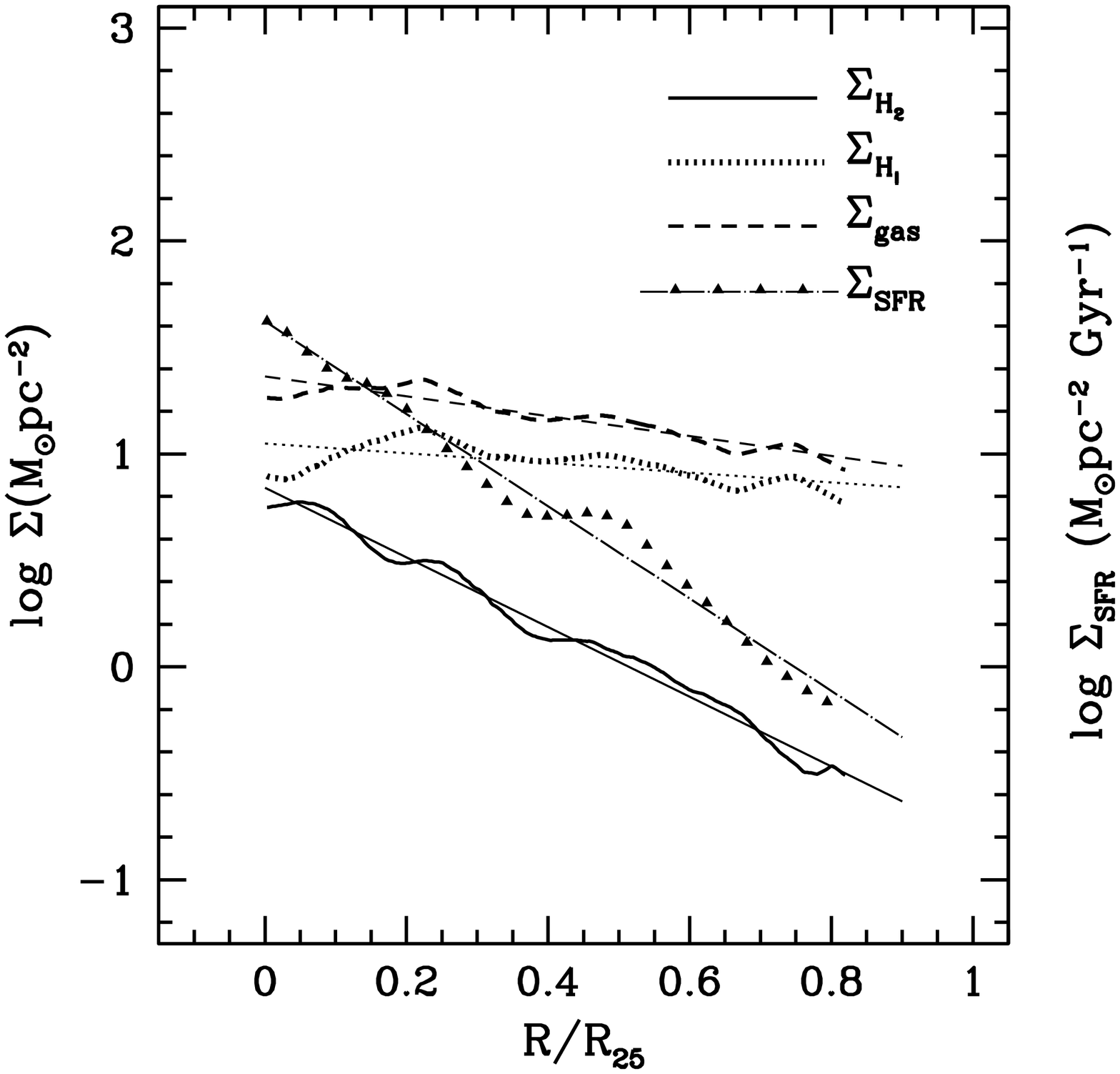}
\caption{M~33, radial distributions. {\it Left panel:} log cluster mass. 
{\it Grey dots:} cluster data;
{\it blue crosses} and {\it blue short-dashed line:} average of 5 most massive 
clusters in bins $4\farcm04$ wide (i.e., 1 kpc, $\approx 0.11 R_{25}$), and weighted fit;
{\it red filled circles} and {\it red solid line:} median of 5 most massive clusters
(i.e., effectively, the third most massive cluster) in the same bins, and fit.  
{\it Right panel:} log surface densities ({\it thick lines}) and
linear fits ({\it thin lines}). {\it Solid:} H$_2$; {\it dotted:} HI; 
{\it dashed:} total gas; {\it solid triangles} and {\it dotted-long-dashed line:} 
SFR. See text for the fitted relation. The $x$- and $y$- axes of the two panels have the same dynamic ranges,
so that the slopes of the fits to the cluster and gas data are
directly comparable.
} 
\label{fm33rad}
\end{figure*}

Results are shown in
Figure~\ref{fm33rad}. In the left panel, we plot log$_{10}$ of cluster mass vs.\ galactocentric
radius in units of $R_{25}$, for the objects in the \citet{shar11} sample
with mass $M_{\rm clust} \geq 10^3 M_\odot$.\footnote{Three clusters at galactocentric 
distances 1.07 $R_{25}$, 1.20 $R_{25}$, and 1.28 $R_{25}$, with masses, respectively, 
$1.6 \times 10^3\ M_\odot$, $5 \times\ 10^4 M_\odot$, and $3.2\ \times 10^3 M_\odot$ are out of the figure, to the right.} 
We constructed 7 bins covering the range with both neutral and 
molecular gas data, i.e., between the galactic center and 0.8 $R_{25}$, each one 4$\farcm$04 
(1 kpc, $\approx 0.11\ R_{25}$) wide. 
The mean of the five most massive
clusters in each bin, and a weighted\footnote{
The mean of each bin $i$ is weighted by $w_i = 1/\sigma_i^2$, where $\sigma_i$ is its dispersion.} 
linear fit to the mean are
shown, respectively, as blue crosses and a blue short-dashed line; the error bars are
those of the mean. 
The median of the five most massive clusters in the same bins
(or the third most massive cluster in each bin), and a linear fit to the
median are displayed as
red filled circles and a solid line; the error bars represent the
interquartile range. Since the uncertainty in the measurement 
of any individual mass is typically a factor of 2-3, fits to the mean and the
median of the 5 most massive clusters in each bin should in general be more robust
indicators of any existing trend (in every bin there are always more than 
5 clusters more massive than the completeness limit). Using this mean or median is even more necessary in this case,
since we intend to compare cluster masses, not with the gas surface
densities at their position, but with azimuthal averages at their
galactocentric distance. The RMS azimuthal variations are always below 5\% of
the CO and mid-IR emission, and below 2\% of the HI and HI+CO emission.

From the weighted fit to the mean,
we obtain: 
\begin{equation}
{\rm log}_{10}\ M_{\rm mean}/M_\odot = (-2.2 \pm 0.2) R/R_{25} 
+ (5.2 \pm 0.1); 
\label{radialmeans}
\end{equation}
\noindent the fit to the median yields:
\begin{equation}
{\rm log}_{10}\ M_{\rm 3rd}/M_\odot = (-2.2 \pm 0.2) R/R_{25} 
+ (5.1 \pm 0.1). 
\label{radialmedians}
\end{equation}

We note that there is an extremely massive cluster at
$\approx 0.85\ R_{25}$, almost at the edge of the
available H$_2$ data and beyond the published 
radial profiles of $\Sigma_{\rm H2}$ and 
$\Sigma_{\rm SFR}$. This cluster, the most massive in the galaxy 
with $M_{\rm clust} \ga 10^6 M_\odot$,
sits over one of the brightest spots in the CO J=1-0
map, while there is very little
emission detected from the rest of the galaxy at the
same galactocentric distance \citep[see][their Figure 2.]{heye04}. 
The region is also quite bright at 21 cm, 
and is the brightest location between 24 and 170 $\mu$m
\citep{heye04,hipp03,taba07}. Hence, 
the azimuthally averaged emission of the corresponding annulus 
would not be 
representative of the star-forming conditions at the 
location of the cluster. This cluster has, instead, 
formed in a local major instability in the interstellar medium.

The SFR and gas surface densities 
\citep{heye04}, and our linear
fits to the data are shown in the right panel of Figure~\ref{fm33rad}. The x-axes of both
panels in the figure are the same, whereas the y-axes have the same dynamic ranges, so
that the slopes of fits to gas surface densities and cluster masses are
directly comparable. 

The fits to the available gas are:
\begin{eqnarray}
{\rm log}_{10}\ \Sigma_{\rm gas} = (-0.47 \pm 0.01) R/R_{25} + (1.363 \pm 0.005);\nonumber\\
{\rm log}_{10}\ \Sigma_{\rm H_2} = (-1.63 \pm 0.01) R/R_{25} + (0.84 \pm 0.01);\nonumber\\
{\rm log}_{10}\ \Sigma_{\rm SFR} = (-2.17 \pm 0.02) R/R_{25} + (1.62 \pm 0.01);\nonumber\\
{\rm log}_{10}\ \Sigma_{\rm H_I} = (-0.23 \pm 0.02) R/R_{25} + (1.05 \pm 0.01).\nonumber\footnotemark
\end{eqnarray}
\footnotetext{
Heyer et al.\ find $\Sigma_{\rm gas} \propto R/R_{25}^{2/3}$, 
but they do not show the HI data they used. 
We have obtained the fits to the gas profiles 
and to the cluster masses in a consistent fashion. 
}  
\noindent
\noindent
For M~33, then, both the mean and the median of the
5 most massive clusters yield: 
$M_{\rm mean,3rd} \propto \Sigma_{\rm gas}^{4.7 \pm 0.4}$;
$M_{\rm mean,3rd} \propto \Sigma_{\rm H_2}^{1.3 \pm 0.1}$;
$M_{\rm mean,3rd} \propto \Sigma_{\rm SFR}^{1.0 \pm 0.1}$.

We discard here the possibility that the change in mean and median 
maximum cluster mass with radius is a statistical  
effect, due to the number of clusters in each
equal sized bin decreasing with increasing radius. If clusters
are drawn purely randomly, or stochastically, from the same mass distribution 
function that declines with mass, 
and that has a constant upper truncation mass $M_{\rm u} > 10^6 M_\odot$, 
the probability of picking a massive cluster decreases with the size of the sample, 
i.e., with the number of clusters in the sample.
We demonstrate in the following that the change in maximum mass is due to physical causes instead.

\begin{figure*}
\includegraphics[angle=-90.,width=1.00\hsize,clip=]{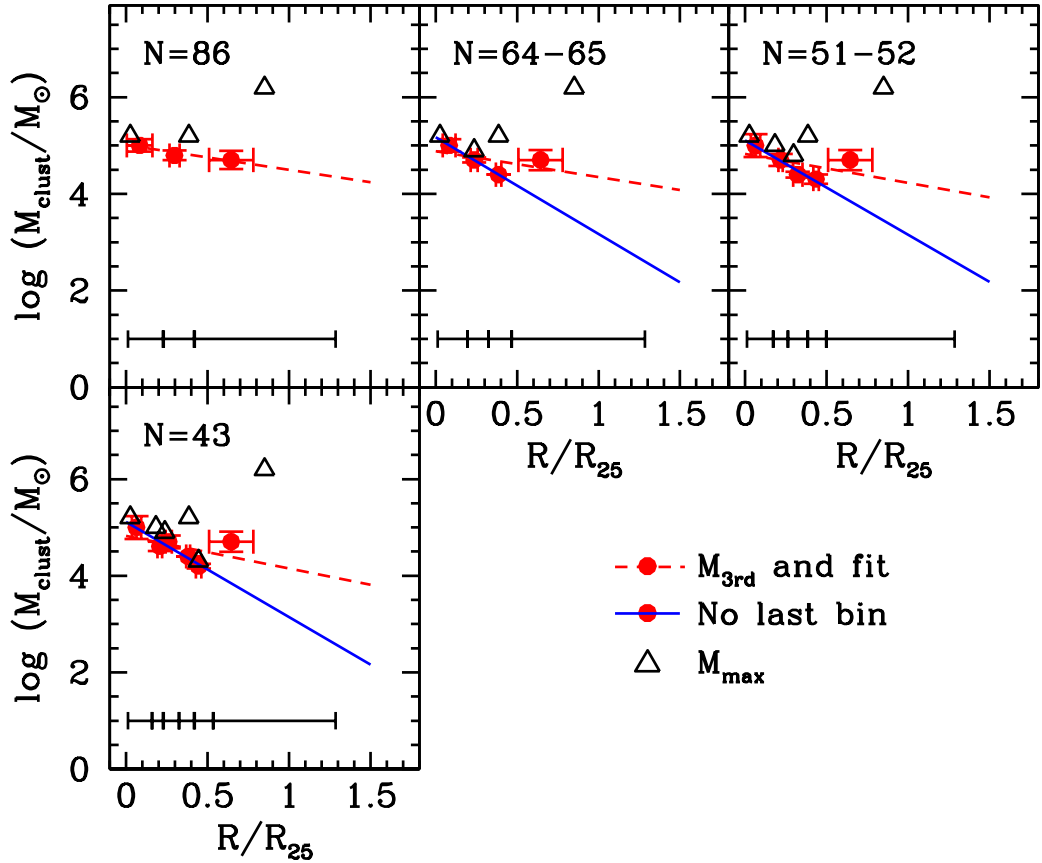}
\caption{M~33, log cluster mass vs.\ radius, fits to bins with equal numbers of clusters. The number of bins
increases from 3 (upper left panel) to 6 (lower left panel); numbers of 
clusters in each bin are indicated. The bar at the bottom
of each panel shows radial ranges of the bins. 
{\it Empty triangles:} most massive cluster in each bin;
{\it filled red circles:} 
median of five most massive clusters (i.e., third most massive cluster) in each bin;
{\it red dashed line:} fit to medians in all bins; {\it blue solid line:} fit to medians
excluding last bin. Fits are performed only for three and more bins. 
}
\label{m33_sos}
\end{figure*}

Figure~\ref{m33_sos} shows fits to the median mass
clusters in bins, each one containing an 
equal number of objects from the subsample of
258 clusters with $M_{\rm clust} \geq 10^3 M_\odot$ detected by 
\citet{shar11} in M~33; the total number
of bins increases from 3 (upper left panel) to 6 (lower left panel).
The number of clusters in each bin is indicated, and the maximum 
and median masses are shown, respectively, with
black empty triangles and red filled circles. 

Two fits to the medians are performed: one including all
the bins in each panel (red dashed line), and one omitting
the last bin (blue solid line); this bin includes a radial range 
for which there is no molecular gas data, and always contains the 
most massive cluster in M~33 (see above), an object whose characteristics
do not seem to correlate with the average star-forming conditions 
at its galactocentric radius. 
The fits have the form:
\begin{equation} 
{\rm log_{10}}\ M_{\rm clust}/M_\odot = (\beta^\prime \pm \sigma_{\beta^\prime}) R/R_{25} + (\alpha^\prime \pm \sigma_{\alpha^\prime}),
\label{fiteq}
\end{equation}
\noindent
and their coefficients and their uncertainties are listed, respectively, in the top
and bottom of Table 1. 
When the last bin is included, the slopes $\beta^\prime$ of these fits 
get slowly steeper, as the number of bins increases (see top of Table 1) and
the relative importance of
the last bin progressively diminishes. If, on the other hand, 
we exclude the last bin, 
we find that the slope of the fits to the median cluster mass 
is $\beta^\prime = -2.0 \pm0.3$ (see Table 1, bottom), 
regardless of the number of bins.

\begin{deluxetable}{rrrrrrr}
\tabletypesize{\small}
\tablecolumns{5}
\tablewidth{15cm}
\tablecaption{Fits to log $M_{\rm 3rd/M_\odot}$ vs.\ $R/R_{25}$, Figure 2 and eq.~3}
\tablehead{
\colhead{$N_b$}&
\colhead{$N_{\rm cl}$}&
\colhead{$\beta^\prime \pm \sigma_{\beta^\prime}$}&
\colhead{$\alpha^\prime \pm \sigma_{\alpha^\prime}$}&
\colhead{RMS}
}
\startdata
 3 & 86&  $(-0.5 \pm 0.2)$ & $(5.01 \pm 0.08)$& 0.04 \\
 4 & 64-65 & $(-0.5 \pm 0.6)$ & $(4.9 \pm 0.2)$& 0.2  \\
 5 & 51-52 & $(-0.6 \pm 0.6)$ & $(4.8 \pm 0.2)$& 0.2 \\
 6 & 43& $(-0.7 \pm 0.6)$ & $(4.8 \pm 0.2)$ & 0.2 \\ \hline
\multicolumn{5}{c}{Without last bin} \\ \hline
 3 & 64-65 & $(-2.00 \pm 0.02)$ & $(5.17 \pm 0.01)$ & 0.003 \\
 4 & 51-52 & $( -2.0 \pm 0.3 )$ & $(5.1 \pm 0.1)$ & 0.05 \\
 5 & 43& $(-2.0 \pm 0.3)$ & $(5.1 \pm 0.1)$ & 0.1  \\ 
\enddata
\tablecomments{Col.\ (1): number of bins. Col.\ (2): 
number of clusters in each bin. Col.\ (3): best-fit slope.
Col.\ (4): best-fit intercept. Col.\ (5): best-fit RMS residual. 
}
\label{tabdat}
\end{deluxetable}

If, instead of the median, we consider the 90$^{\rm th}$ percentile cluster
(i.e., the 9$^{\rm th}$, 6$^{\rm th}$, 5$^{\rm th}$, and
4$^{\rm th}$ most massive cluster, respectively, for 3, 4, 5, and 6
bins), the correlation (including the last bin) is consistent
in all cases with ${\rm log_{10}}\ M_{\rm 90th}/M_\odot = 1.2\pm0.2 R/R_{25} + (4.7 \pm 0.1)$. 
Thus, the relation between cluster mass and
radius is slightly shallower than for $M_{\rm 3rd}$, but
still quite robust and definitely not flat. We note here, though, that for large
samples (for example, of a few hundred clusters) the mass of the 90$^{\rm th}$ percentile
point might already be one or two orders of magnitude below 
that of the most massive clusters, and hence would not be a good 
estimator for this particular problem. 

In order to convince the reader that the intrinsic mass distribution of
clusters changes as a function of radius, we perform a Kolmogorov-Smirnov
(K-S) test on the 6 different radial subsamples presented in the bottom 
left panel of Figure 2. Figure 3 shows the cumulative probability 
distributions and median radii of each subsample, and the $D$ and $P$ values
for every bin pair are given in Table 2 (higher bin number indicates larger
radius). The visual impression, that the inner bins are different
from the outer ones, is confirmed by the K-S statistic:
assuming that sample pairs with $P < 0.05$ are
taken from different distribution functions with
high significance, we conclude that bins 1 and 2, within
$R = 0.22\ R_{25}$, are 
different from bins 5 and 6, beyond
$R = 0.42\ R_{25}$. Bin 3 also 
differs from bin 6, such that the change of the 
cluster mass distribution function appears to be gradual, 
rather than stepwise.  

\begin{figure*}
\includegraphics[angle=0.,width=1.00\hsize,clip=]{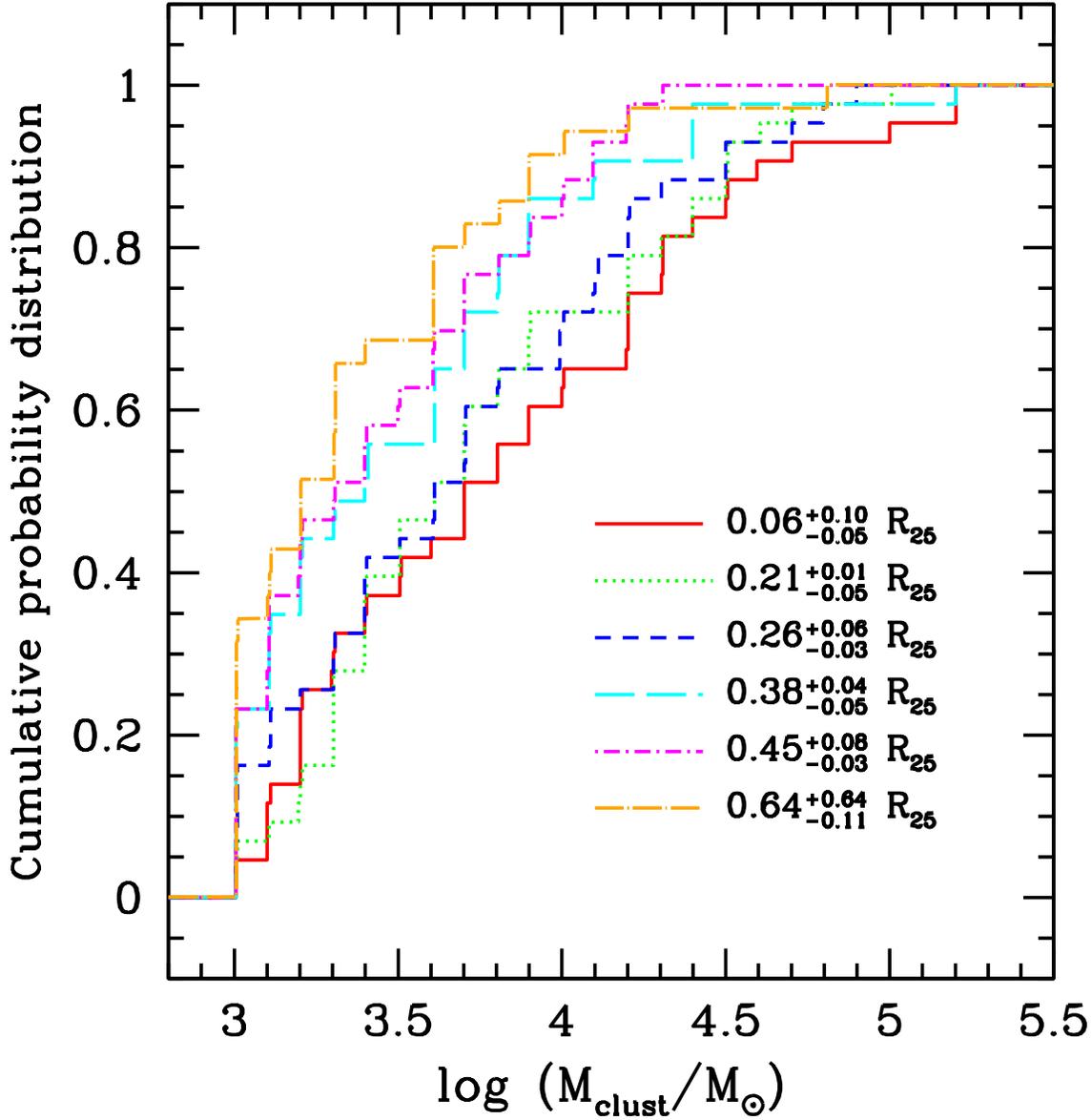}
\caption{K-S test; cumulative probability distributions
of mass for clusters with $M_{\rm clust} \geq 10^3\ M_\odot$ in the six
annuli whose median radii are indicated. {\it Red solid line:} 
bin 1, $R = 0.06^{+0.10}_{-0.05}\ R_{25}$; 
{\it green dotted line:} bin 2, $R = 0.21^{+0.01}_{-0.05}\ R_{25}$; 
{\it blue short dashed line:} bin 3, $R = 0.26^{+0.06}_{-0.03}\ R_{25}$;
{\it cyan long dashed line:} bin 4, $R = 0.38^{+0.04}_{-0.05}\ R_{25}$; 
{\it purple short dashed-dotted line:} bin 5, $ R = 0.45^{+0.08}_{-0.03}\ R_{25}$; 
{\it orange long dashed-dotted line:} bin 6, $R = 0.64^{+0.64}_{-0.11}\ R_{25}$. 
}
\label{m33_ks}
\end{figure*}

\begin{table*}
\center{\sc Table 2\\ K-S test $D$ and $P$ values}\\
\center{}
  \hspace*{1.25cm}
 \begin{minipage}{140mm}
\begin{small}
\begin{tabular}{@{}c|cc|cc|cc|cc|cc@{}}
\hline 
\hline 
\vspace*{-0.021cm}& \vspace*{-0.021cm}& \vspace*{-0.021cm}&\vspace*{-0.021cm}& \vspace*{-0.021cm} &\vspace*{-0.021cm} &\vspace*{-0.021cm} &\vspace*{-0.021cm} &\vspace*{-0.021cm} &\vspace*{-0.021cm} &\vspace*{-0.021cm}  \\
 & $D$ & $P$ & $D$ & $P$ & $D$ & $P$ & $D$ & $P$ & $D$ & $P$  \\ 
\vspace*{-0.026cm}& \vspace*{-0.026cm}& \vspace*{-0.026cm}&\vspace*{-0.026cm}& \vspace*{-0.026cm} &\vspace*{-0.026cm} &\vspace*{-0.026cm} &\vspace*{-0.026cm} &\vspace*{-0.026cm} &\vspace*{-0.026cm} &\vspace*{-0.026cm} \\ \hline
\multicolumn{1}{c}{Bin}&\multicolumn{2}{|c|}{2}&\multicolumn{2}{|c|}{3}&\multicolumn{2}{|c|}{4}&\multicolumn{2}{|c|}{5}&\multicolumn{2}{|c}{6} \\ \hline
1 & 0.56 & 0.9168 & 0.78 & 0.5799 & 1.22 & 0.1005 & 1.45 & 0.0307 & 1.63  & 0.0096   \\ \hline
2 &      &        & 0.67 & 0.7651 & 1.33 & 0.0569 & 1.45 & 0.0307 & 1.72  & 0.0052   \\ \hline
3 &      &        &      &        & 1.00 & 0.2694 & 1.11 & 0.1687 & 1.63  & 0.0096   \\ \hline
4 &      &        &      &        &      &        & 0.56 & 0.9168 & 1.10  & 0.1797   \\ \hline
5 &      &        &      &        &      &        &      &        & 0.69  & 0.7271  \\
\hline
\tablecomments{\small $D$ and $P$ values for bin pairs. The cell in the intersection
of a row and a column contains the $D$ and $P$
parameters, respectively, of the comparison between the two bins indicated in the
corresponding row and column. If $P < 0.05$, the null hypothesis that the 
clusters in the two bins are taken from the same mass distribution function is rejected.}
\end{tabular}
\end{small}
\end{minipage}
\label{kstab}
\end{table*}

These results rule out the size of sample effect, according to which
no correlation ($\beta^\prime = 0$) is expected
between maximum cluster mass and galactic radius. 
From the ratio $\sigma_{\beta^\prime}/\beta^\prime$, it 
follows that the falsification of stochastic
sampling is at a confidence level of at least 7$\sigma$.

\section{Conclusions}

We have analyzed the relationship between maximum cluster mass and  gas 
surface density in M~33, in order to explore the suggestion that 
maximum cluster mass is determined by physical processes, e.g., 
the equilibrium pressure between cluster forming cores and the 
ambient interstellar medium \citep{lars02,bill02}, and 
since the existence of such a relationship can reconcile, via the IGIMF theory, SFR measurements 
derived, respectively, from H$\alpha$ and FUV emission in galaxy disks.  

To this end, we have used published gas data of M~33 \citep{corb03,heye04}, and a catalog
of more than 600 young star clusters in its disk, also from the
literature \citep{shar11}. 
Because often the most massive cluster in a range of
galactocentric distances is formed under conditions that are not average
for the annulus, we find that it is best, in the present annular averaging
approach, to use the median of the five most massive
clusters (i.e, the third most massive cluster) in each
bin for this kind of analysis.   

We have compared radial distributions, and have found that
$M_{\rm 3rd} \propto \Sigma_{\rm H_2}^{1.3 \pm 0.1}$, 
while $M_{\rm 3rd} \propto \Sigma_{\rm SFR}^{1.0 \pm 0.1}$ (in a range consistent with the
expectations from pressure equilibrium considerations). 
On the other hand, $M_{\rm max} \propto \Sigma_{\rm gas}^{4.7 \pm 0.4}$,
steeper than needed to explain the H$\alpha$ cut-off in galaxy disks.
Both this correlation and the steeper than average star formation law
might be related to the shallowness of the HI profile.\footnote{
It is very likely that M~33 has interacted with M~31 in the past. 
The radial HI density profile of M~33 may have been changed away from
an exponential disk as a result, and hence may not reflect at present 
the physically relevant conditions for star formation in a 
virialized self-regulated galactic disk in equilibrium. Another example of a 
transitory state of the HI gas is the Magellanic Stream, parts of which
will most likely be re-accreted onto the Large Magellanic Cloud
once it orbits to a larger distance from the MW.}

In order to test whether the trend of $M_{\rm max}$ (or its proxy,
$M_{\rm 3rd}$), with galactocentric radius 
is consistent with random sampling from the cluster mass function, we have also measured the
radial distribution of maximum mass in 3 to 6 bins with an equal number of 
clusters in each bin. After accounting for the presence of the
most massive cluster in the galaxy at $\approx 0.85 R_{25}$, whose formation
environment is certainly not represented by the average conditions 
across the galaxy at its galactocentric distance, and for the
lack of gas data beyond this same radius,
we find exactly the same results as before, regardless of the width of the
bins. A K-S test on the mass distributions in these bins suggests
that the two bins closest to the galaxy center are different from
the two most external ones, i.e., that the mass distribution
function changes with radius. 

The significant decrease of ${\rm log}\ M_{\rm max}$ with radial distance in M~33,
as a power law with index $\beta^\prime \approx -2.0 \pm 0.3$, and despite there being
the same number of clusters per radial bin, rules out random sampling 
with extremely high confidence. This one galaxy, thus, falsifies the
hypothesis that the most massive cluster masses scale with the size of
the sample. Instead, the range of star cluster masses is driven by 
environmental physics.
Indeed, the available data for M~33 suggest log$_{10}\ M_{\rm 3rd} \propto\ {\rm log}_{10}\Sigma_{\rm H_2}
\propto\ {\rm log}_{10}\Sigma_{\rm SFR}$. This, however, may be merely a (trivial) 
self-consistent result because, after all, stars may form from molecular clouds 
in a free-fall timescale and subsequently destroy the clouds \citep{hart01}. The non-trivial challenge remains to
answer how the interstellar medium, and thus mostly the HI gas, arranges itself to
form molecular clouds.  

\acknowledgments
We thank Mark Heyer, who provided us with measurements of the
azimuthal variations in the gas data, and the anonymous referee, 
for his/her positive and helpful feedback. RAGL acknowledges
support from DGAPA, UNAM.


\begin{thebibliography}{}


\bibitem[Bell 
\& Kennicutt(2001)]{bell01} Bell, E.~F., \& Kennicutt, R.~C., Jr.\ 2001, \apj, 548, 681 

\bibitem[Billett et al.(2002)]{bill02} Billett, O.~H., Hunter, 
D.~A., \& Elmegreen, B.~G.\ 2002, \aj, 123, 1454 

\bibitem[Boissier et al.(2007)]{bois07} Boissier, S., Gil de 
Paz, A., Boselli, A., et al.\ 2007, \apjs, 173, 524 

\bibitem[Boselli et al.(2009)]{bose09} Boselli, A., Boissier, 
S., Cortese, L., et al.\ 2009, \apj, 706, 1527 






\bibitem[Corbelli(2003)]{corb03} Corbelli, E.\ 2003, \mnras, 342, 199


\bibitem[Corbelli 
\& Schneider(1997)]{corb97} Corbelli, E., \& Schneider, S.~E.\ 1997, \apj, 479, 244 

\bibitem[Demleitner et al.(2001)]{deml01} Demleitner, M., 
Accomazzi, A., Eichhorn, G., et al.\ 2001, Astronomical Data Analysis 
Software and Systems X, 238, 321 

\bibitem[Elmegreen(1983)]{elme83} Elmegreen, B.~G. 1983, MNRAS, 2003, 1011


\bibitem[Freedman et al.(1991)]{free91} Freedman, W.~L., 
Wilson, C.~D., \& Madore, B.~F.\ 1991, \apj, 372, 455 


\bibitem[Gil de Paz et al.(2007)]{gild07} Gil de Paz, A., 
Boissier, S., Madore, B.~F., et al.\ 2007, \apjs, 173, 185 


\bibitem[Greenawalt et al.(1998)]{gree98} Greenawalt, B., 
Walterbos, R.~A.~M., Thilker, D., \& Hoopes, C.~G.\ 1998, \apj, 506, 135 


\bibitem[Hartmann et al.(2001)]{hart01} Hartmann, L., 
Ballesteros-Paredes, J., \& Bergin, E.~A.\ 2001, \apj, 562, 852 

\bibitem[Heyer et al.(2004)]{heye04} Heyer, M.~H., Corbelli, 
E., Schneider, S.~E., \& Young, J.~S.\ 2004, \apj, 602, 723 

\bibitem[Hippelein et 
al.(2003)]{hipp03} Hippelein, H., Haas, M., Tuffs, R.~J., et al.\ 2003, \aap, 407, 137 

\bibitem[Hoopes 
\& Walterbos(2000)]{hoop00} Hoopes, C.~G., \& Walterbos, R.~A.~M.\ 2000, \apj, 541, 597 



\bibitem[Kennicutt(1998)]{kenn98} Kennicutt, R.~C., Jr.\ 1998, \araa, 36, 189


\bibitem[Kroupa
\& Weidner(2003)]{krou03} Kroupa, P., \& Weidner, C.\ 2003, \apj,598, 1076

\bibitem[Larsen(2002)]{lars02} Larsen, S.~S.\ 2002, \aj, 124, 
1393 

\bibitem[Lada 
\& Lada(2003)]{lada03} Lada, C.~J., \& Lada, E.~A.\ 2003, \araa, 41, 57 



\bibitem[Lee et al.(2009)]{lee09} Lee, J.~C., Gil de Paz, A., 
Tremonti, C., et al.\ 2009, \apj, 706, 599 

\bibitem[Leitherer et al.(1999)]{leit99} Leitherer, C., 
Schaerer, D., Goldader, J.~D., et al.\ 1999, \apjs, 123, 3 



\bibitem[Marks 
\& Kroupa(2012)]{mark12} Marks, M., \& Kroupa, P.\ 2012, arXiv:1205.1508 

\bibitem[Maschberger 
\& Kroupa(2007)]{masc07} Maschberger, T., \& Kroupa, P.\ 2007, \mnras, 379, 34 



\bibitem[McConnachie et al.(2010)]{mcco10} McConnachie, A.~W., 
Ferguson, A.~M.~N., Irwin, M.~J., et al.\ 2010, \apj, 723, 1038 

\bibitem[Meurer et al.(2009)]{meur09} Meurer, G.~R., Wong, 
O.~I., Kim, J.~H., et al.\ 2009, \apj, 695, 765 



\bibitem[Pflamm-Altenburg 
\& Kroupa(2008)]{pfla08} Pflamm-Altenburg, J., \& Kroupa, P.\ 2008, \nat, 455, 641 

\bibitem[Pflamm-Altenburg et al.(2009)]{pfla09} 
Pflamm-Altenburg, J., Weidner, C., \& Kroupa, P.\ 2009, \mnras, 395, 394 



\bibitem[Sharma et 
al.(2011)]{shar11} Sharma, S., Corbelli, E., Giovanardi, C., Hunt, L.~K., \& Palla, F.\ 2011, \aap, 534, A96 

\bibitem[Schmidt(1959)]{schm59} Schmidt, M.\ 1959, \apj, 129,
243

\bibitem[Schmidt(1963)]{schm63} \samename 1963, \apj, 137,




\bibitem[Silk(1997)]{silk97} Silk, J.\ 1997, ApJ, 481, 703

\bibitem[Sullivan et al.(2000)]{sull00} Sullivan, M., Treyer, 
M.~A., Ellis, R.~S., et al.\ 2000, \mnras, 312, 442 

\bibitem[Tabatabaei et 
al.(2007)]{taba07} Tabatabaei, F.~S., Beck, R., Krause, M., et al.\ 2007, \aap, 466, 509 



\bibitem[V{\'a}zquez 
\& Leitherer(2005)]{vazq05} V{\'a}zquez, G.~A., \& Leitherer, C.\ 2005, \apj, 621, 695 

\bibitem[Verley et 
al.(2007)]{verl07} Verley, S., Hunt, L.~K., Corbelli, E., \& Giovanardi, C.\ 2007, \aap, 476, 1161 

\bibitem[Weidner \& Kroupa(2004)] {weid04b} Weidner, C., \& Kroupa, P.\ 2004,
\mnras, 348, 187

\bibitem[Weidner
\& Kroupa(2005)]{weid05} Weidner, C., \& Kroupa, P.\ 2005, \apj,
625, 754

\bibitem[Weidner 
\& Kroupa(2006)]{weid06} Weidner, C., \& Kroupa, P.\ 2006, \mnras, 365, 1333 

\bibitem[Weidner et al.(2010)]{weid10} Weidner, C., Kroupa, 
P., \& Bonnell, I.~A.~D.\ 2010, \mnras, 401, 275 

\bibitem[Weidner et al.(2004)]{weid04a} Weidner, C., Kroupa,
P., \& Larsen, S.~S.\ 2004, \mnras, 350, 1503


\end{thebibliography}
\end{document}